# Line and Point Defects in MoSe$_2$ Bilayer Studied by Scanning Tunneling Microscopy and Spectroscopy

Hongjun Liu[1*], Hao Zheng[1,2,3*], Fang Yang[2], Lu Jiao[1], Jinglei Chen[1], Wingkin Ho[1], Chunlei Gao[2,3], Jinfeng Jia[2,3] and Maohai Xie[1†]

[1]*Physics Department, The University of Hong Kong, Pokfulam Road, Hong Kong*

[2]*Key Laboratory of Artificial Structures and Quantum Control (Ministry of Education), Department of Physics and Astronomy, Shanghai Jiaotong University, 800 Dongchuan Road, Shanghai 200240, China*

[3]*Collaborative Innovation Center of Advanced Microstructures, Department of Physics and Astronomy, Shanghai Jiao Tong University, Shanghai 200240, P. R. China*

## Abstract:

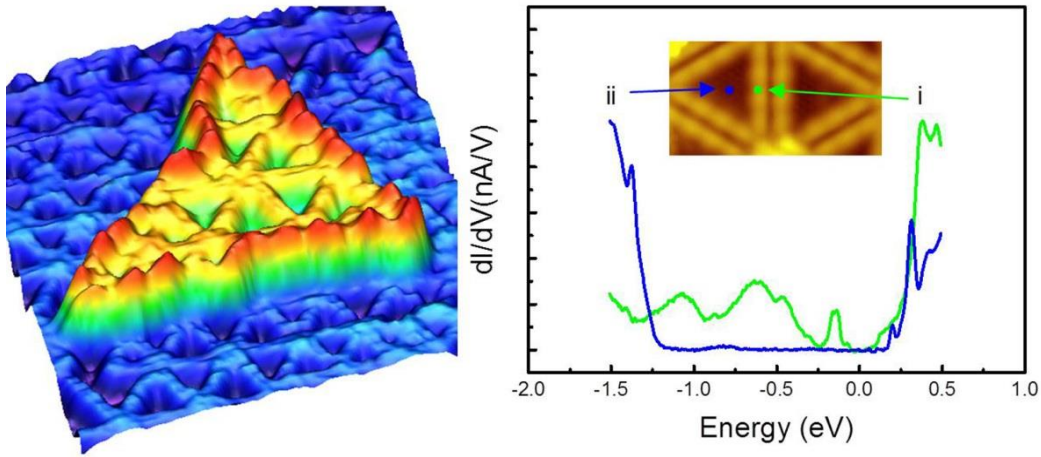

Bilayer (BL) MoSe$_2$ films grown by molecular-beam epitaxy (MBE) are studied by scanning tunneling microscopy and spectroscopy (STM/S). Similar to monolayer (ML) films, networks of inversion domain boundary (DB) defects are observed both in the top and bottom layers of BL MoSe$_2$, and often they are seen spatially correlated such that one is on top of the other. There are also isolated ones in the bottom layer without companion in the top-layer and are detected by STM/S through quantum tunneling of the defect states through the barrier of the MoSe$_2$ ML. Comparing the DB states in BL MoSe$_2$ with that of ML film reveals some common features as well as differences. Quantum confinement of the defect states is indicated. Point defects in BL MoSe$_2$ are also observed by STM/S, where ionization of the donor defect by the tip-induced electric field is evidenced. These results are of great fundamental interests as well as practical relevance of devices made of MoSe$_2$ ultrathin layers.

[*]H.L. and H.Z. contributed equally to this work.
[†]mhxie@hku.hk

Transition metal dichalcogenides (TMDs) represent a family of the layered compounds that are attracting extensive research attention lately.[1-3] Ultrathin TMD films (e.g., $MoS_2$, $WS_2$, $MoSe_2$, and $WSe_2$) possess sizable quasi-particle bandgaps as well as exotic spin-valley coupled electronic structures,[4-13] which hold the promise for ultrathin electronic, optoelectronic, spin- and valley-tronic device applications. As two-dimensional (2D) materials, their high surface-to-volume ratio makes them particularly sensitive to changes in their surroundings and the inertness of the van der Waals (vdW) surfaces makes defect or edge states in the 2D material more prominent. Line and point defects in ultrathin TMD films will thus bring strong effects in the electronic, optical and catalytic properties of the materials.[14-21] The past studies of the TMDs for their intrinsic and defect induced properties have primarily focused on monolayer (ML) films. On the other hand, TMD bilayer (BL) and heterostructures represent another important aspect of the 2D material research, revealing distinctly different properties from ML films.[22-32] Investigations of defects in BL TMDs will thus be of great fundamental interests as well as practical relevance.

In this work, we employ low temperature (LT) scanning tunneling microscopy and spectroscopy (STM/S) to probe the line and point defects in epitaxial $MoSe_2$ BL grown by molecular-beam epitaxy (MBE). We reveal the tunneling effect of inversion domain boundary (DB) defects through the barrier of $MoSe_2$ ML. The mid-gap, one-dimensional (1D), states are confined within finite lengths of the defects, so the quantum well states (QWS) are observed in the LT-STS measurements. By comparing the STS spectra of the DB defects in BL and ML films, we note some common features as well as differences, indicating an effect of interlayer coupling. Electric field induced ionization of a point defect in BL $MoSe_2$ is also followed by the STM/S measurements, which would unveil the Coulomb potential and dielectric properties of films. The latter represent key parameters of ultrathin TMDs dictating the fundamental properties such as strong exciton binding energy of the materials.

**Results and Discussions**

BL MoSe$_2$ thin films were grown on highly ordered pyrolytic graphite (HOPG) substrates in a customized Omicron MBE system at 400 ℃ under excess Se condition. LT-STM/S measurements of the samples were done in a Unisoku system at 4 K, where the constant current mode was adopted for the STM measurements and the lock-in technique was employed for the STS experiments.

(i) *Inversion Domain boundary defects in BL MoSe$_2$*

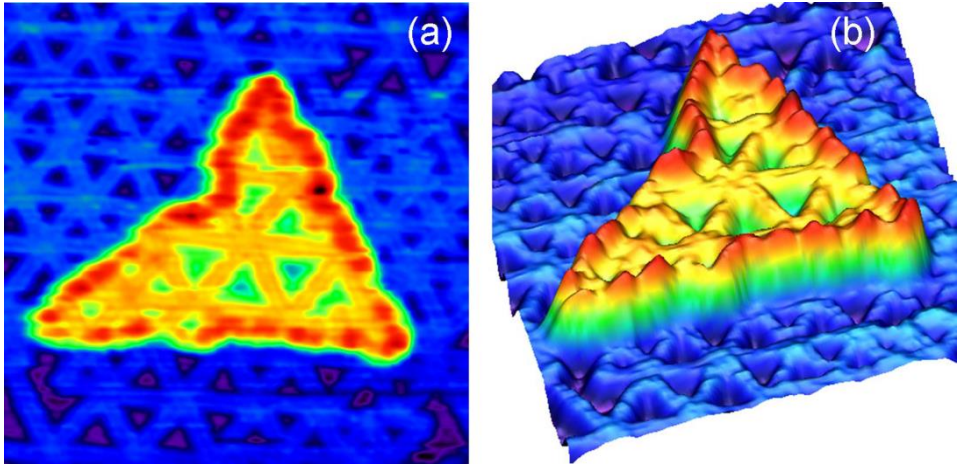

**Fig. 1** Top (**a**) and perspective view (**b**) of an as-grown MoSe$_2$ film surface measured by room-temperature STM (size: 50×50 nm$^2$, sample bias: -1.0 V). The blue and orange colors encode the first (bottom) and second (top) MoSe$_2$ layer respectively in this nominal 1.4 MLs coverage sample.

**Fig. 1** presents a STM topographic image of an as-grown MoSe$_2$ sample in both the top and perspective views, showing both ML and BL MoSe$_2$ films. The sample has the nominal 1.4 MLs coverage, so it has a complete first (or bottom) layer and a partial coverage of the second (top) layer of MoSe$_2$ as displayed in different colors. A striking feature seen on the surface of MoSe$_2$ is the network of bright rims forming the wagon-wheel like patterns. As has been discussed in detail in Ref. [16], similar bright rims in ML MoSe$_2$ reflect additional electronic states introduced by the DB defects. The DB formation in MBE-grown MoSe$_2$ has been further found to closely depend on the MBE conditions[33] and a recent report revealed a possible vacancy-induced formation mechanism of such defects in the TMDs.[34] Statistical analysis of the patterns in BL samples reveals the sizes of the wagon-wheels are slightly larger than that in ML film. This may partly be due to a variation of the MBE condition in the course of the growth experiment or else related to the different strains in the top *versus* bottom layers of epitaxial $MoSe_2$. Notably, the DB defects in the two layers appear spatially correlated, *i.e.*, they often align vertically and so the ones in the second layer are dominantly on top of those in the first (bottom) layer. However, because of a lower density of the defects in the top layer, there are also isolated DBs in the first layer showing no companion in the top-layer. These defects are detected by STM due to quantum tunneling effect through the "barrier" of the top $MoSe_2$ ML, which will be elaborated further in a later discussion. For now, we concentrate on the morphological and electronic structures of the DB defects in BL $MoSe_2$. As in ML films, each DB manifests in STM images as symmetrical twin lines, along which intensities are undulated. In an early study,[16] we attributed the twin-line structure to the very spatial distribution of the density of states (DOS) of the defects. For the undulated intensities, on the other hand, we associated them to the QWS as well as an effect of the moiré superlattice potential. The defects in BL $MoSe_2$ have given rise to similar mid-gap electronic states as in ML $MoSe_2$, which are evidenced in the dI/dV spectrum of **Fig. 2(i)** taken on one of the bright twin-lines of a defect. This is compared to **Fig. 2(ii)** taken from a defect-free region of BL $MoSe_2$, which shows semiconductor characteristics with a bandgap of ~ 1.7 eV. For comparison, we also present a spectrum of the DB defect in $MoSe_2$ ML (*i.e.*, spectrum of **Fig. 2(iii)**).

The ML and BL TMDs are known to have distinctly different properties.[25,35,36] For example, ML $MoSe_2$ is a direct gap semiconductor with the bandgap of ~ 2.25 eV while a BL $MoSe_2$ is an indirect gap semiconductor of a much reduced bandgap.[36,37] Studies have also shown that for pristine TMD bilayers, coupling of electronic states between the two layers leads to upward(downward) shift of the Γ(Q) states in the valence(conduction) bands containing predominantly the *p*-orbital components of the chalcogen atoms.[36] The K valley states, which are mainly of the *d*-orbitals of the metal, show little change upon film thickness increase from ML to BL. Consequently, the materials transfer from direct to indirect gap semiconductors with increasing film thickness. For the DB defects, the induced mid-gap states expectedly suffer from the same effects and will exhibit some common and different features in the STS spectra between ML and BL $MoSe_2$. Particularly, because the defects in the two layers of the BL $MoSe_2$ film are spatially correlated, the close proximity of the defects will lead to coupling. Comparing (i) and (iii), one observes groups of conductance peaks, such as those circled by the solid and dashed black lines, to appear similar between ML and BL films. They also show approximately the same energy separation despite the shifts of the absolute energy positions between ML and BL spectra. The latter may be attributed partly to the band bending effect or the difference in electron affinity between ML and BL films,[38] or else to an effect of coupling of the defect states in the BL film. In spectrum (i), one further observes additional peaks (*e.g.*, arrow pointed ones) that are not present in the ML spectrum, signifying the dependence of the defect states on film thickness.

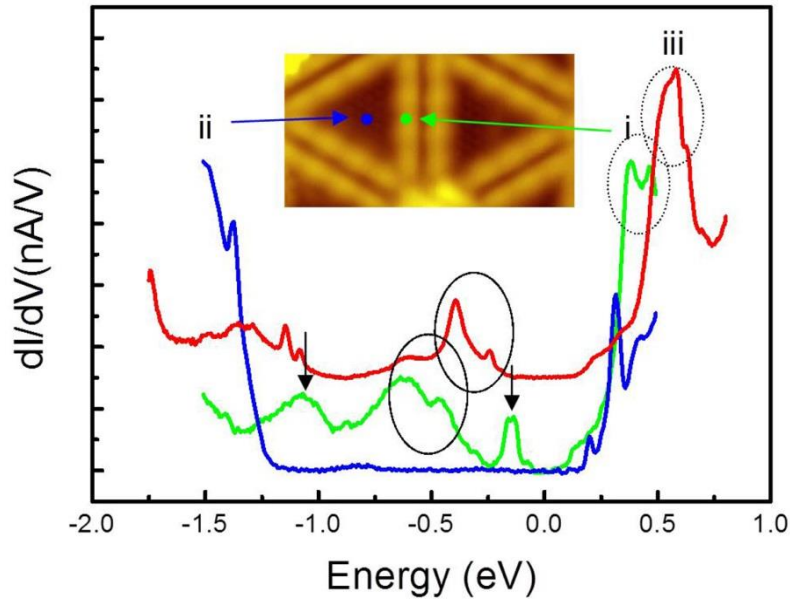

**Fig. 2** STS spectra taken (**i**) on one of the bright twin lines of the DB defects and (**ii**) in the defect-free region of BL $MoSe_2$. (**iii**) STS spectrum for a DB defect in ML $MoSe_2$. The inset marks the positions in a STM micrograph (size: 9×5 nm$^2$) where spectra (i) and (ii) were obtained.

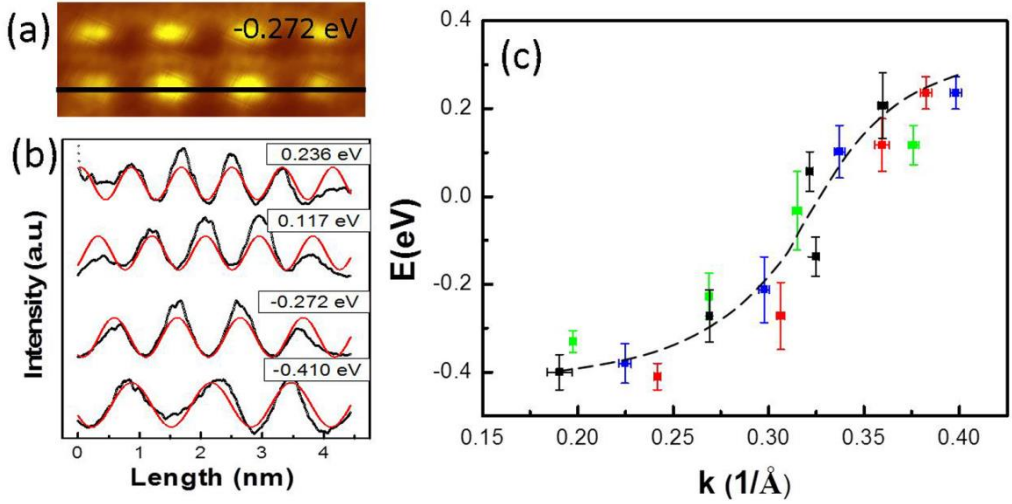

**Fig. 3** (**a**) STS map of a domain boundary defect measured at the energy as labelled. The black line marks the position where the intensity profiles in (b) are extracted from. (**b**) Intensity profiles of the defect in (a) at different energies. The solid lines represent the fitting results by the function $I = A\sin^2(kx + \varphi) + D$, where $A$, $k$, $\varphi$ and $D$ are fitting parameters. (**c**) The fitted wavevector $k$ for a few different defects (coded by colors) and at different energies. The dashed line is drawn to guide the reader's eye, showing a dispersive $E \sim k$ relation.

As in ML MoSe$_2$, intensity undulations of the defect states reflect the QWS along the 1D defect. For purposes of more quantitative analyses, we take a series of STS measurements at different energies and **Fig. 3a** presents an example of the STS images for a particularly defect. We then extract the intensity profiles (*I*) along the defect and at different energies. The results are presented in Fig. 3b, in which the superimposed solid lines are least-square fittings of the data by the function $I = A\sin^2(kx + \varphi) + D$, where $A$, $k$, $\varphi$, and $D$ are constants. The derived wavevectors (*k*) are summarized in **Fig. 3c** for a few defects of different lengths, revealing a dispersive $E \sim k$ relation. Notably, for a given defect we find the wavevectors satisfy the relation $\frac{k'}{k} = \frac{N}{M}$ (*N* and *M* are integers), confirming the QWS. Different from ML MoSe$_2$,[16] where states with wavevectors larger than 0.32 Å$^{-1}$ were not observed, in the BL film, states of $k > 0.32$ Å$^{-1}$ are recorded. As remarked in Ref. [16], the periodic moiré superlattice potential in the MoSe$_2$/graphene system leads to the band-folding effect, and the dominant scattering events confined within the reduced Brillouin zone have an upper limit for the scattered wavevectors as dictated by the period of the moiré superlattice potential and is ~ 0.32 Å$^{-1}$. In BL MoSe$_2$, the same moiré potential may have weakened significantly for electrons in the top layer, so Bragg reflection by the moiré superlattice potential is not realized in BL MoSe$_2$. Therefore we observe waves of $k > 0.32$ Å$^{-1}$ in Fig. 3b.

(ii) *Tunneling of the DB states through MoSe$_2$*

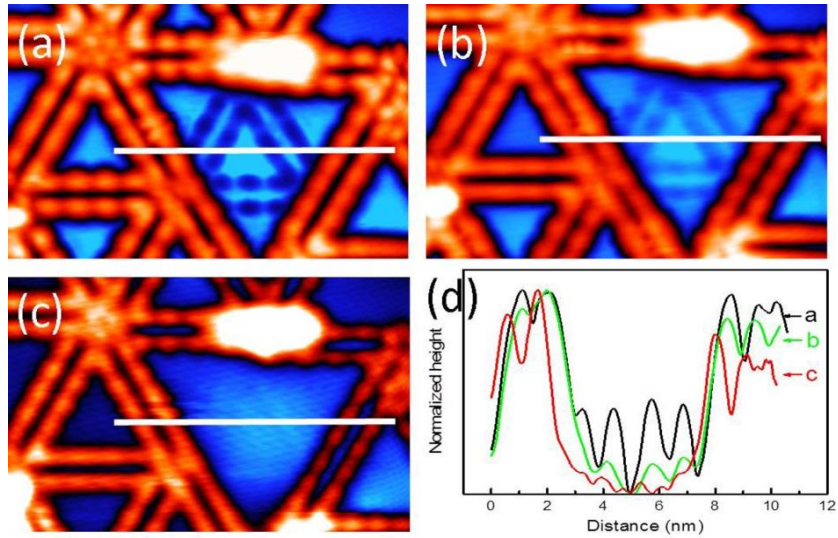

**Fig. 4** STM images acquired over BL MoSe$_2$ at (**a**) -1.29 V, (**b**) -1.97 V and (**c**) -2.39 V, respectively, of the same area (size: 14 nm x 9 nm). The line defects are displayed in orange. In (a, b), weak contrast features are seen within a defect-free region of the top layer, which are assigned to be the 'buried' line defects in the bottom layer. (**d**) Line profiles taken along the white lines drawn in (a-c). They are normalized per the maximum intensity.

Visualization by the STM/S of the DB defects in the bottom-layer *via* quantum tunneling through the top MoSe$_2$ ML is remarkable, which deserves further elaborations. **Figs. 4(a-c)** present examples of the STM images at different bias conditions. Besides the bright rims representing the DB defects in the top layer, there are also lines of weaker contrasts but otherwise the same morphology as the bright rims. On the first sight, this morphological feature of Fig. 4 resembles that of the lateral heterojunction between graphene and BN,[39,40] where different STM/S contrasts were revealed between the boron- and nitrogen-terminated edges of BN due to varying DOS at the two junctions. Here, on the other hand, we assert that the weaker lines in Figs. 4a and 4b are the DBs in the *bottom layer* of the BL film, which are seen in STM micrographs due to tunneling of the defect states through the top $MoSe_2$ ML. Comparing the three images in Fig. 4, one observes that the darker rims change their contrasts with the bias and become increasingly invisible upon decreasing the sample bias. **Fig. 4d** presents line profiles measured from the three images on the same position, elucidating the diminishing contrast with decreasing energy. By referring to the STS spectrum of Fig. 2(ii), we see that Fig. 4a corresponds to an energy that is just inside the bandgap of BL $MoSe_2$. Under such a condition, electrons of the defect states in the bottom layer tunnel into the STM tip through the "energy gap" of top-layer $MoSe_2$, and so they are visible in STM micrographs. This is similar to a recent report of tunneling by the Dirac states in graphene through ML $WSe_2$ in an STS experiment.[41] Contrasting to STS, which exhibits local DOS at a particular energy, STM images of Fig. 4 represent integrated DOS over the energy range from the Fermi level to the applied sample bias. Therefore, the defects can still be visible at the bias outside the bandgap (*e.g.*, Fig. 4b) (note, however, that there are resonant states, which will also contribute to the STM contrast). On the other hand, because the tunneling matrix is dominated by states at the energy of bias, the contrast of the defects in STM become increasingly weaker as the bias is further shifted from the band gap edge to deeper in the valance band (Fig. 4c).

Visualization of the DB defects in the bottom layer through the top $MoSe_2$ ML suggests significant out-of-plane momenta of the defect states. They could be the $d_{z^2}$ component of the metal atoms or the *p*-states of Se. Considering the distances of the two atoms from the STM tip, the *p*-orbitals of Se can be more relevant. Such out-of-plane components of the defect states will make them strongly coupled between the DBs in the top and bottom layers of BL MoSe$_2$ if are vertically correlated. This is consistent with an early discussion about the new feature in the STS spectrum of a BL MoSe$_2$ than ML film.

(iii) *Ionization of point defect in BL MoSe$_2$ by tip-induced electric field*

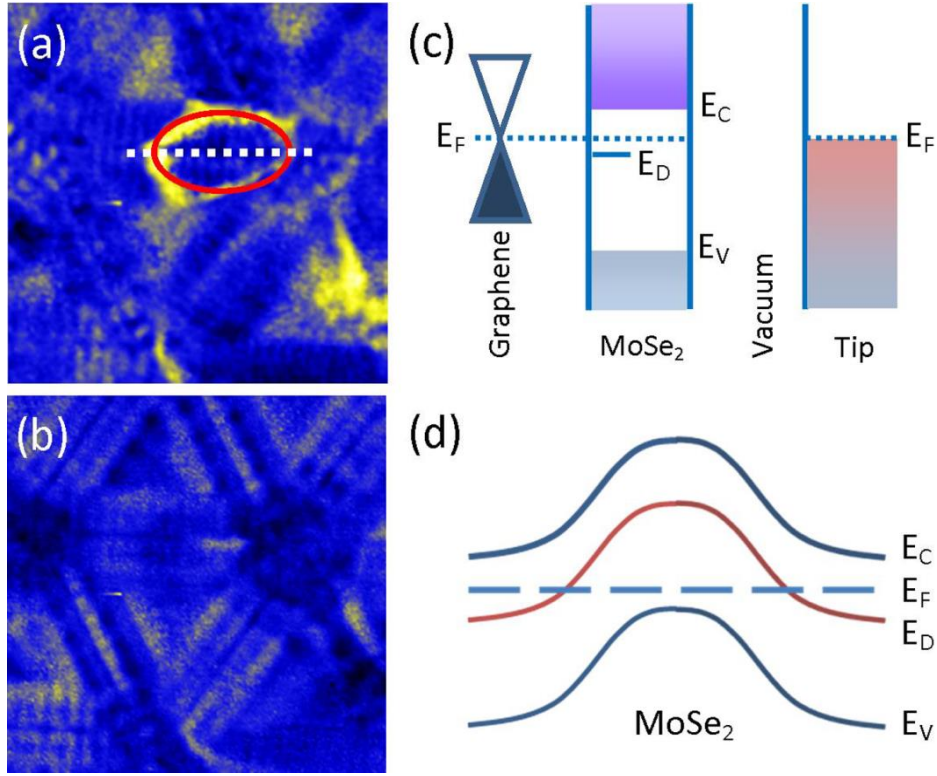

**Fig. 5** (**a, b**) dI/dV maps (size: 9 × 9 nm$^2$) acquired at 0.56 V and 0.38 V, respectively, revealing the presence and absence of an elliptical ionization ring as highlight by the solid red line in (a). The dotted horizontal line in (a) marks the position along which Fig. 6a below is obtained. (**c**) Energy diagram of the tip–MoSe$_2$/graphite tunneling junction at zero bias. (**d**) Schematic illustration of tip induced band bending in the plane of MoSe$_2$. In (c) and (d), E$_c$ and E$_v$ mark the conduction band minimum and valence band maximum, respectively. E$_F$ is the Fermi level and E$_D$ is the defect state with a constant binding energy relative to E$_C$.

In addition to the line defects visualized by STM/S, point defects are also observed by STM/S through the ionization process by the tip-induced electric field.

For example, **Fig. 5a** shows a STS map of a MoSe$_2$ BL, in which an elliptical bright feature is clearly discernable. As one decreases the bias of the measurements, the size of the ellipse shrinks (see **Supplementary S1**) and becomes invisible below 0.38 V (**Fig. 5b**). This morphological feature and its bias-dependence signify an ionization process of a donor defect in MoSe$_2$.[42-45]

Low electron density on the surface of MoSe$_2$ cannot shield the tip induced electric field during STM measurements, so the field effect inside the semiconductor MoSe$_2$ must be taken into account. At a sample bias above the "flat-band" condition, the penetration of the electric field in ultrathin MoSe$_2$ layer lifts the energy bands upward, inducing a transition of the charge state of a donor defect at a sufficiently high voltage (see **Fig. 5c**). This, in turn, leads to an enhancement in the tunneling current and manifests in the STS map by a bright ring such as in Fig. 5a.[42-45] Obviously, the electric field is the strongest direct underneath the tip, so the band shift is the most prominent at zero lateral distance from the tip. Moving away from the tip position, the field strength decays monotonically with the lateral distance, causing varying degrees of band lifting (refer to a schematic diagram of **Fig. 5d**). At a given bias voltage, there exists a specific distance of the tip from the defect such that the defect level $E_D$ intersects with the Fermi energy $E_F$. At such distances, transition of the charge state of the defect activates, which gives rise to enhanced STS contrast. For a point defect imaged by a symmetric tip, a circular ring will show up with the diameter reflecting the ionization energy of the defect as well as its depth in location from the sample surface.[42,46-48] On the other hand, a non-ideal shape of the STM tip apex would generate deviations, but the above interpretation of the phenomenon remains valid. The experiment of Fig. 5 provides a direct evidence of electric field induced ionization of a donor defect.

**Fig. 6a** is a color-coded position-dependent dI/dV diagram along the marked dotted line in Fig. 5a and at varying energies. The nearly parabolic curve conforms to a donor defect in MoSe$_2$. In the diagram, distance zero (*i.e.*, the position where the parabola minimum lies) refers to the lateral position where the defect is. Moving away from the defect, higher voltages are required to ionize the defect. The defect appears singly charged, as we do not observe double rings over wide energy range scans. **Fig. 6b** shows two individual STS spectra taken at different lateral positions relative to the defect (marked in Fig. 6a), from which one observes clearly the ionization peak at energies above the conduction band edge, e.g., at about 0.6 eV in the top curve taken from inside the ionization ring of the dI/dV map. It is absent in the spectra taken from outside of the ionization ring (*e.g*., the bottom curve). This reaffirms the electric field effect inducing the ionization of donor defect in MoSe$_2$.

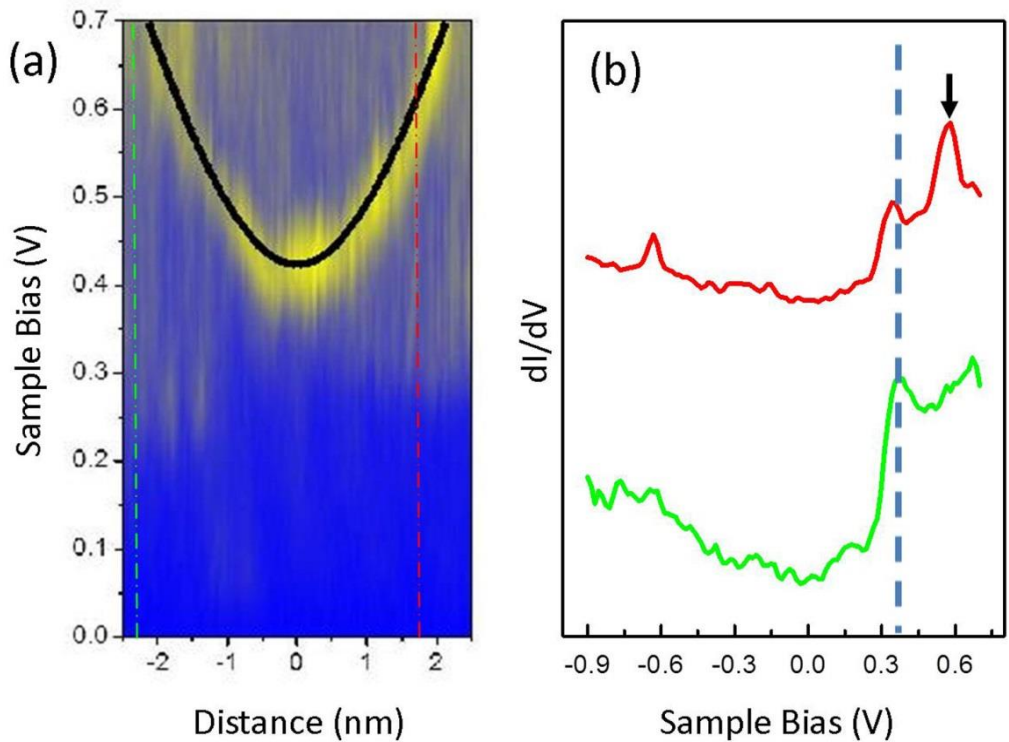

**Fig. 6** (**a**) Color-coded diagram showing position-dependent dI/dV spectra (it contains a total of 108 individual spectra). The yellow parabola reveals the spatial evolution of the ionization peaks. The black line represents a simulated result (see text). The vertical dash-dotted lines mark the positions where the STS spectra in (b) were derived. (**b**) Individual dI/dV spectra taken at different distances from the defect as indicated by the colored vertical dash-dotted lines in (a). The arrow points to the ionization peak in the top(red) curve. The dashed line marks the conduction band edge.

In principle, one can perform simulations of the energy-dependent ionization ring or the position-dependent dI/dV diagram (*i.e.*, the parabola) to derive the binding energy and/or other properties of the defect and the host material. There are, however, complications due to a few unknown parameters, such as the STM tip characteristics and the contact potential, which would affect the fitting results.[42] Nevertheless, based on appropriate assumptions of such parameters, one may still provide order-of-magnitude estimates of the defect binding energy and/or the dielectric constant of the film. For ultrathin TMDs, one of the important material properties is the dielectric constants ($\varepsilon_r$). The much reduced $\varepsilon_r$ in ML and BL TMDs than that of the bulk crystal contributes to the enhanced exciton binding energies, for example.[49] It is thus quite appealing to provide independent experimental evaluations of such a parameter than the exciton binding energy measurements by optical methods. Coulomb potential profile associated with an ionized donor defect in a film will contain such information, so the result of Fig. 6 can be used to extract such a quantity. Following Ref. [42], we performed simulations of the result of Fig. 6a and the black line represents one of the simulated results assuming the dielectric constant of BL MoSe$_2$ is $\varepsilon_r$ ~ 4.6 as suggested by theory.[49] The resulted binding energy of the defect is ~ 240 meV. If one uses a larger $\varepsilon_r$, such as that of the bulk MoSe$_2$ (*i.e.*, ~ 10.2),[49] the fitted result appears also good (see **Supplementary S2**), and the derived binding energy will be lower at ~ 130 meV. To discriminate one from the other, one needs a priori knowledge about the defect. For a donor defect in MBE-grown MoSe$_2$, the most relevant candidate includes Se-vacancy, metal-adatom, interstitial or antisite. Unfortunately, we do not have enough data to ascertain the identity of the defect. Should we accept the theoretical dielectric constant ($\varepsilon_r$ ~ 4.6), the derived binding energy of ~ 240 meV may actually help to specify the very identify of the defect in MBE-grown MoSe$_2$, a key property in its own right in semiconductor physics and devices.

**Conclusion**

BL MoSe$_2$ films grown by MBE are investigated by STM/S, revealing both DB and point donor defects. The DB defects give rise to mid-gap states and their quantum confinement leads to undulated intensities in the STM/S micrographs. By comparing ML with BL MoSe$_2$, a possible coupling of the defect states between the top and bottom layers of the BL film is indicated. Tunneling of the DB states through MoSe$_2$ ML is also visualized in STM measurements. Electric field induces ionization of a point defect in BL MoSe$_2$ and manifests by a bright ionization ring in the STS map. Simulations of the energy dependent ionization rings and/or position-dependent conductance spectra may lead to derivations of such important properties as the binding energy of the defect and the dielectric property of the host material. These STM/S studies of the defect states, including direct visualization of the tunneling effect and the electric field induced ionization processes, are imperative and important to supplement the transport and optical studies of the devices made of the TMD films.

**Methods**

MoSe$_2$ films were deposited on freshly cleaved HOPG from elemental molybdenum (Mo) and selenium (Se) sources in an e-beam evaporator and a dual-filament Knudsen cell, respectively, using a customized Omicron MBE reactor. The flux of the metal source was calibrated by the built-in flux monitor in an e-beam evaporator and that of Se was estimated by the beam-equivalent pressure (BEP) measured using a beam flux monitor at the sample position. The flux ratio between Mo and Se was 1:15 and the film deposition rate was 0.5 MLs per hour (*i.e.*, ~ 0.3 nm/hr). The sample temperature during growth was 400 ℃. Reflection high-energy electron diffraction (RHEED) operated at 10 keV was employed to monitor the growing surfaces and the persistent streaky RHEED patterns signified 2D layer-by-layer growth mode of MoSe$_2$ on HOPG. After a preset coverage of the deposit was grown, the source fluxes were stopped by closing the mechanical shutters and in the meantime the sample was cooled naturally to room-temperature (RT) for subsequent RT-STM measurements using an Omicron VTSTM system. For some, the samples were capped by amorphous Se layers by depositing Se at RT, which could be signified by the change of the streaky RHEED patterns of MoSe$_2$ to diffusive ones. Such samples were transported to a separate Unisoku LT-STM system for STM/S characterizations at 4 K. Prior to the LT-STM/S measurements, however, clean and smooth MoSe$_2$ surfaces were recovered by thermal desorption of the Se capping layers. The latter were verified by the reappearance of the sharp streaky RHEED patterns as well as the observation of smooth surface morphology of the samples by STM. For both RT and LT-STM experiments, the constant current mode was adopted. For the STS measurements, the lock-in technique was employed using the modulation voltage of 15 mV and frequency of 985 Hz. The differential conductance (dI/dV) spectra were taken at each point of an image with same starting voltage and current and the same lock-in parameters. The dI/dV map was acquired by plotting the dI/dV data at a particular energy for every point of the image.

**Conflict of Interest**

The authors declare no competing financial interest.

**Acknowledgements**

We appreciate the comments and suggestions from W. Yao. This work is supported by a Collaborative Research Fund (HKU9/CRF/13G) sponsored by the Research Grant Council (RGC), Hong Kong Special Administrative Region. H.L. and M.X. acknowledge the support of the internal grant of HKU. The work is also supported from the SRFDP and RGC ERG Joint Research Scheme of Hong Kong RGC and the Ministry of Education of China (No. M-HKU709/12). The work in SJTU was also supported by the MOST (2013CB921902, 012CB927401) and the NSFC (11227404, 11374206) of China. C. G. acknowledges the support from the NSFC (11304014) of China.

**References**

1. Wang, Q. H.; Kalantar-Zadeh, K.; Kis, A.; Coleman, J. N.; Strano, M. S. Electronics and Optoelectronics of Two-dimensional Transition Metal Dichalcogenides. *Nat. Nanotechnol.* 2012, 7, 699.
2. Mak, K. F.; Lee, C.; Hone, J.; Shan, J.; Heinz, T. F. Atomically Thin $MoS_2$: A New Direct-Gap Semiconductor. *Phys. Rev. Lett.* 2010, 105, 136805.
3. Splendiani, A.; Sun, L.; Zhang, Y.; Li, T.; Kim, J.; Chim, C.-Y.; Galli, G.; Wang, F. Emerging Photoluminescence in Monolayer $MoS_2$. *Nano Lett.* 2010, 10, 1271.
4. Xiao, D.; Liu, G.-B.; Feng, W.; Xu, X.; Yao, W. Coupled Spin and Valley Physics in Monolayers of $MoS_2$ and Other Group-VI Dichalcogenides. *Phys. Rev. Lett.* 2012, 108, 196802.
5. Mak, K. F.; He, K.; Shan, J.; Heinz, T. F. Control of Valley Polarization in Monolayer $MoS_2$ by Optical Helicity. *Nat. Nanotechnol.* 2012, 7, 494.
6. Zeng, H.; Dai, J.; Yao, W.; Xiao, D.; Cui, X. Valley Polarization in $MoS_2$ Monolayers by Optical Pumping. *Nat. Nanotechnol.* 2012, 7, 490.
7. Cao, T.; Wang, G.; Han, W. P.; Ye, H. Q.; Zhu, C. R.; Shi, J. R.; Niu, Q.; Tan, P. H.; Wang, E.; Liu, B. L.; Feng, J. Valley-selective Circular Dichroism of Monolayer Molybdenum Disulphide. *Nat. Commun.* 2012, 3, 887.
8. Jones, A. M.; Yu, H.; Ghimire, N. J.; Wu, S.; Aivazian, G.; Ross, J. S.; Zhao, B.; Yan, J.; Mandrus, D. G.; Xiao, D.; Yao, W.; Xu, X. Optical Generation of Excitonic Valley Coherence in Monolayer $WSe_2$. *Nat. Nanotechnol.* 2013, 8, 634.
9. Ross, J. S.; Wu, S.; Yu, H.; Ghimire, N. J.; Jones, A. M.; Aivazian, G.; Yan, J.; Mandrus, D. G.; Xiao, D.; Yao, W.; Xu, X. Electrical Control of Neutral and Charged Excitons in a Monolayer Semiconductor. *Nat. Commun.* 2013, 4, 1474.
10. Radisavljevic, B.; Radenovic, A.; J. Brivio; Giacometti, V.; Kis, A. Single-layer $MoS_2$ transistors. *Nat. Nanotechnol.* 2011, 6, 147.
11. Zhang, Y.; Chang, T.-R.; Zhou, B.; Cui, Y.-T.; Yan, H.; Liu, Z.; Schmitt, F.; Lee, J.; Moore, R.; Chen, Y. *et al*. Direct Observation of the Transition from Indirect to Direct Bandgap in Atomically Thin Epitaxial $MoSe_2$. *Nat. Nanotechnol.* 2014, 9, 111.
12. Alidoust, N.; Bian, G.; Xu, S.-Y.; Sankar, R.; Neupane, M.; Liu, C.; Belopolski, I.; Qu, D.-X.; Denlinger, J. D.; Chou, F.-C.; Hasan, M. Z. Observation of Monolayer Valence Band Spin-orbit Effect and Induced Quantum Well States in $MoX_2$. *Nat. Commun.* 2014, 5, 4673.
13. Yin, Z.; Li, H.; Li, H.; Jiang, L.; Shi, Y.; Sun, Y.; Lu, G.; Zhang, Q.; Chen, X.; Zhang, H. Single-Layer $MoS_2$ Phototransistors. *ACS Nano* 2011, 6, 74.
14. van der Zande, A. M.; Huang, P. Y.; Chenet, D. A.; Berkelbach, T. C.; You, Y.; Lee, G.-H.; Heinz, T. F.; Reichman, D. R.; Muller, D. A.; Hone, J. C. Grains and Grain Boundaries in Highly Crystalline Monolayer Molybdenum Disulphide. *Nat. Mater.* 2013, 12, 554.
15. Gutiérrez, H. R.; Perea-López, N.; Elías, A. L.; Berkdemir, A.; Wang, B.; Lv, R.; López-Urías, F.; Crespi, V. H.; Terrones, H.; Terrones, M. Extraordinary Room-Temperature Photoluminescence in Triangular $WS_2$ Monolayers. *Nano Lett.* 2013, 13, 3447.
16. Liu, H.; Jiao, L.; Yang, F.; Cai, Y.; Wu, X.; Ho, W.; Gao, C.; Jia, J.; Wang, N.; Fan, H.; Yao, W.; Xie, M. Dense Network of One-Dimensional Midgap Metallic Modes in Monolayer $MoSe_2$ and Their Spatial Undulations. *Phys. Rev. Lett.* 2014, 113, 066105.
17. Tongay, S.; Suh, J.; Ataca, C.; Fan, W.; Luce, A.; Kang, J. S.; Liu, J.; Ko, C.; Raghunathanan, R.; Zhou, J. *et al.* Defects Activated Photoluminescence in Two-dimensional Semiconductors: Interplay between Bound, Charged, and Free excitons. *Sci. Rep.* 2013, 3, 2657.

18. Ghorbani-Asl, M.; Enyashin, A. N.; Kuc, A.; Seifert, G.; Heine, T. Defect-induced Conductivity Anisotropy in $MoS_2$ Monolayers. *Phys. Rev. B* 2013, 88, 245440.

19. Ma, Y.; Dai, Y.; Guo, M.; Niu, C.; Lu, J.; Huang, B. Electronic and Magnetic Properties of Perfect, Vacancy-doped, and Nonmetal Adsorbed $MoSe_2$, $MoTe_2$ and $WS_2$ Monolayers. *Phys. Chem. Chem. Phys.* 2011, 13, 15546.

20. Besenbacher, F.; Brorson, M.; Clausen, B. S.; Helveg, S.; Hinnemann, B.; Kibsgaard, J.; Lauritsen, J. V.; Moses, P. G.; Nørskov, J. K.; Topsøe, H. Recent STM, DFT and HAADF-STEM Studies of Sulfide-based Hydrotreating Catalysts: Insight Into Mechanistic, Structural and Particle Size Dffects. *Catal. Today* 2008, 130, 86.

21. Lehtinen, O.; Komsa, H.-P.; Pulkin, A.; Whitwick, M. B.; Chen, M.-W.; Lehnert, T.; Mohn, M. J.; Yazyev, O. V.; Kis, A.; Kaiser, U.; Krasheninnikov, A. V. Atomic Scale Microstructure and Properties of Se-Deficient Two-Dimensional $MoSe_2$. *ACS Nano* 2015, 9, 3274.

22. Jones, A. M.; Yu, H.; Ross, J. S.; Klement, P.; Ghimire, N. J.; Yan, J.; Mandrus, D. G.; Yao, W.; Xu, X. Spin-layer Locking Effects in Optical Orientation of Exciton Spin in Bilayer $WSe_2$. *Nat. Phys.* 2014, 10, 130.

23. Wu, S. F.; Ross, J. S.; Liu, G. B.; Aivazian, G.; Jones, A.; Fei, Z. Y.; Zhu, W. G.; Xiao, D.; Yao, W.; Cobden, D.; Xu, X. D. Electrical Tuning of Valley Magnetic Moment through Symmetry Control in Bilayer $MoS_2$. *Nat. Phys.* 2013, 9, 149.

24. Wang, H.; Yu, L.; Lee, Y.-H.; Shi, Y.; Hsu, A.; Chin, M. L.; Li, L.-J.; Dubey, M.; Kong, J.; Palacios, T. Integrated Circuits Based on Bilayer $MoS_2$ Transistors. *Nano Lett.* 2012, 12, 4674.

25. He, J.; Hummer, K.; Franchini, C. Stacking Effects on the Electronic and Optical Properties of Bilayer Transition Metal Dichalcogenides $MoS_2$, $MoSe_2$, $WS_2$, and $WSe_2$. *Phys. Rev. B* 2014, 89, 075409.

26. Zhu, B.; Zeng, H.; Dai, J.; Gong, Z.; Cui, X. Anomalously Robust Valley Polarization and Valley Coherence in Bilayer $WS_2$. *Proc. Natl. Acad. Sci.* 2014, 111, 11606.

27. Gong, Z.; Liu, G.-B.; Yu, H.; Xiao, D.; Cui, X.; Xu, X.; Yao, W. Magnetoelectric Effects and Valley-controlled Spin Quantum Gates in Transition Metal Dichalcogenide Bilayers. *Nat. Commun.* 2013, 4, 2053.

28. Geim, A. K.; Grigorieva, I. V. Van der Waals Heterostructures. *Nature* 2013, 499, 419.

29. Fang, H.; Battaglia, C.; Carraro, C.; Nemsak, S.; Ozdol, B.; Kang, J. S.; Bechtel, H. A.; Desai, S. B.; Kronast, F.; Unal, A. A. *et al.* A. Strong Interlayer Coupling in van der Waals Heterostructures Built from Single-layer Chalcogenides. *Proc. Natl. Acad. Sci.* 2014, 111, 6198.

30. Gong, Y.; Lin, J.; Wang, X.; Shi, G.; Lei, S.; Lin, Z.; Zou, X.; Ye, G.; Vajtai, R.; Yakobson, B. I. *et al.* Vertical and In-plane Heterostructures from $WS_2/MoS_2$ Monolayers. *Nat. Mater.* 2014, 13, 1135.

31. Huang, C.; Wu, S.; Sanchez, A. M.; Peters, J. J. P.; Beanland, R.; Ross, J. S.; Rivera, P.; Yao, W.; Cobden, D. H.; Xu, X. Lateral Heterojunctions within Monolayer $MoSe_2$–$WSe_2$ Semiconductors. *Nat. Mater.* 2014, 13, 1096.

32. Duan, X.; Wang, C.; Shaw, J. C.; Cheng, R.; Chen, Y.; Li, H.; Wu, X.; Tang, Y.; Zhang, Q.; Pan, A.; Jiang, J.; Yu, R.; Huang, Y.; Duan, X. Lateral Epitaxial Growth of Two-dimensional Layered Semiconductor Heterojunctions. *Nat. Nanotechnol.* 2014, 9, 1024.

33. Jiao, L.; Liu, H. J.; Chen, J. L.; Yi, Y.; Chen, W. G.; Cai, Y.; Wang, J. N.; Dai, X. Q.; Wang, N.; Ho, W. K.; Xie, M. H. Molecular-Beam Epitaxy of Monolayer $MoSe_2$: Growth Characteristics and Domain Boundary Formation. *New. J. Phys.* 2015, 17, 053023.

34. Lin, J.; Pantelides, S.T.; Zhou, W. Vacancy-Induced Formation and Growth of Inversion Domains in Transition-Metal Dichalcogenide Monolayer. *ACS Nano 2015*, 9, 5189

35. Debbichi, L.; Eriksson, O.; Lebègue, S. Electronic Structure of Two-dimensional Transition Metal Dichalcogenide Bilayers from *ab initio* Theory. *Phys. Rev. B* 2014, 89, 205311.

36. Liu, G.-B.; Xiao D.; Yao Y. G.; Xu X. D.; Yao W. Electronic Structures and Theoretical Modelling of Two-dimensional Group-VIB Transition Metal Dichalcogenides. *Chem. Soc. Rev.* 2015, 44, 2643.

37. Zhang, C.; Chen, Y.; Johnson, A.; Li, M.-Y.; Huang, J.-K.; Li, L.-J.; Shih, C.-K. Measuring Critical Point Energies in Transition Metal Dichalcogenides. 2014, arXiv:1412.8487v1.

38. Liu, H.; Jiao, L.; xie, L.; Yang, F.; Chen, J.; Ho, W.; Gao, C.; Jia, J.; Cui, X.; Xie, M. Molecular-Beam Epitaxy of Monolayer and Bilayer $WSe_2$: A Scanning Tunneling Microscopy/Spectroscopy Study and Deduction of Exciton Binding Energy. *2D Mater.* 2015, in press.

39. Park, J.; Lee, J.; Liu, L.; Clark, K. W.; Durand, C.; Park, C.; Sumpter, B. G.; Baddorf, A. P.; Mohsin, A.; Yoon, M.; Gu, G.; Li, A.-P. Spatially Resolved One-dimensional Boundary States in Graphene–Hexagonal Boron Nitride Planar Heterostructures. *Nat. Commun.* 2014, 5 : 5403.

40. Liu, L.; Park, J.; Siegel, D. A.; McCarty, K. F.; Clark, K. W.; Deng, W.; Basile, L.; Idrobo, J. C.; Li, A.-P.; Gu, G. Heteroepitaxial Growth of Two-Dimensional Hexagonal Boron Nitride Templated by Graphene Edges. *Science* 2014, 343, 163.

41. Chiu, M.-H.; Zhang, C.; Shiu, H. W.; Chuu, C.-P.; Chen, C.-H.; Chang, C.-Y. S.; Chen, C.-H.; Chou, M.-Y.; Shih, C.-K.; Li, L.-J. Determination of Band Alignment in Transition Metal Dichalcogenides Heterojunctions. 2014, arXiv:1406.5137v3.

42. Teichmann, K.; Wenderoth, M.; Loth, S.; Ulbrich, R. G.; Garleff, J. K.; Wijnheijmer, A. P.; Koenraad, P. M. Controlled Charge Switching on a Single Donor with a Scanning Tunneling Microscope. *Phys. Rev. Lett.* 2008, 101, 076103.

43. Lee, D.-H.; Gupta, J. A. Tunable Control over the Ionization State of Single Mn Acceptors in GaAs with Defect-Induced Band Bending. *Nano Lett.* 2011, 11, 2004.

44. Zheng, H.; Kröger, J.; Berndt, R. Spectroscopy of Single Donors at ZnO(0001) Surfaces. *Phys. Rev. Lett.* 2012, 108, 076801.

45. Song, C.-L.; Sun, B.; Wang, Y.-L.; Jiang, Y.-P.; Wang, L.; He, K.; Chen, X.; Zhang, P.; Ma, X.-C.; Xue, Q.-K. Charge-Transfer-Induced Cesium Superlattices on Graphene. *Phys. Rev. Lett.* 2012, 108, 156803.

46. Wijnheijmer, A. P.; Garleff, J. K.; Teichmann, K.; Wenderoth, M.; Loth, S.; Ulbrich, R. G.; Maksym, P. A.; Roy, M.; Koenraad, P. M. Enhanced Donor Binding Energy Close to a Semiconductor Surface. *Phys. Rev. Lett.* 2009, 102, 166101.

47. Zheng, H.; Weismann, A.; Berndt, R. Manipulation of Subsurface Donors in ZnO. *Phys. Rev. Lett.* 2013, 110, 226101.

48. Zheng, H.; Weismann, A.; Berndt, R. Tuning the Electron Transport at Single Donors in Zinc Oxide with a Scanning Tunnelling Microscope. *Nat. Commun.* 2014, 5, 2992.

49. Kumar, A.; Ahluwalia, P. K. Tunable Dielectric Response of Transition Metals Dichalcogenides $MX_2$ (M=Mo, W; X=S, Se, Te): Effect of Quantum Confinement. *Physica B* 2012, 407, 4627.

# Line and Point Defects in MoSe$_2$ Bilayer Studied by Scanning Tunneling Microscopy and Spectroscopy

Hongjun Liu[1*], Hao Zheng[1,2,3*], Fang Yang[2], Lu Jiao[1], Jinglei Chen[1], Wingkin Ho[1], Chunlei Gao[2,3], Jinfeng Jia[2,3] and Maohai Xie[1†]

## Supplementary information

**Figure S1** Sets of voltage-depended dI/dV maps showing the evolution of the ionization rings with energy (image size: 9 nm ×9nm, tunneling current: 224 pA).

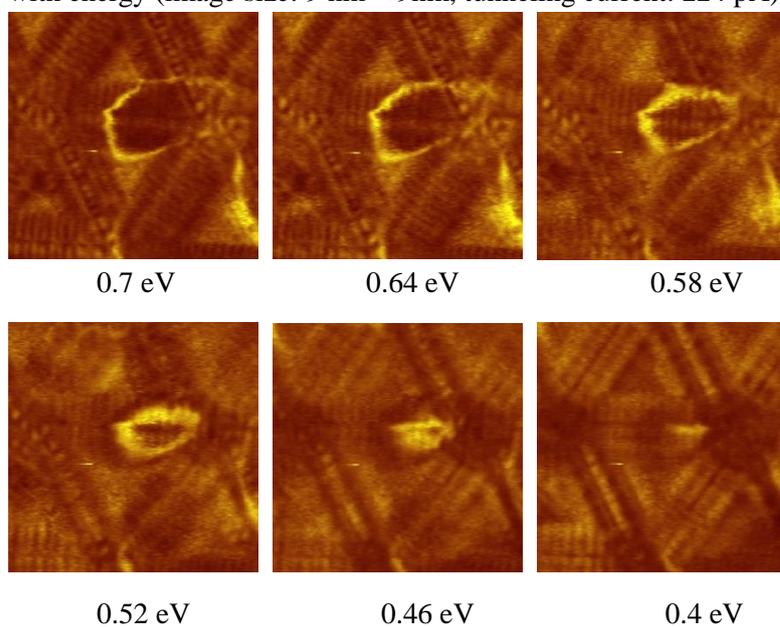

0.7 eV      0.64 eV      0.58 eV

0.52 eV      0.46 eV      0.4 eV

**Figure S2** Simulations of the ionization of a point defect. The black lines display the simulated curves assuming a point defect being buried one layer below surface with the dielectric constant 4.6 (a) and 10.2 (b). The derived binding energies are marked against each line. Due to many critical parameters, e.g. the tip radius, tip-sample distance, contact potential, are missing, the values of the binding energy only present rough estimates. Nevertheless, the simulated curves fit well with the experimental data, lending support of the model and underlying physics of the phenomenon.

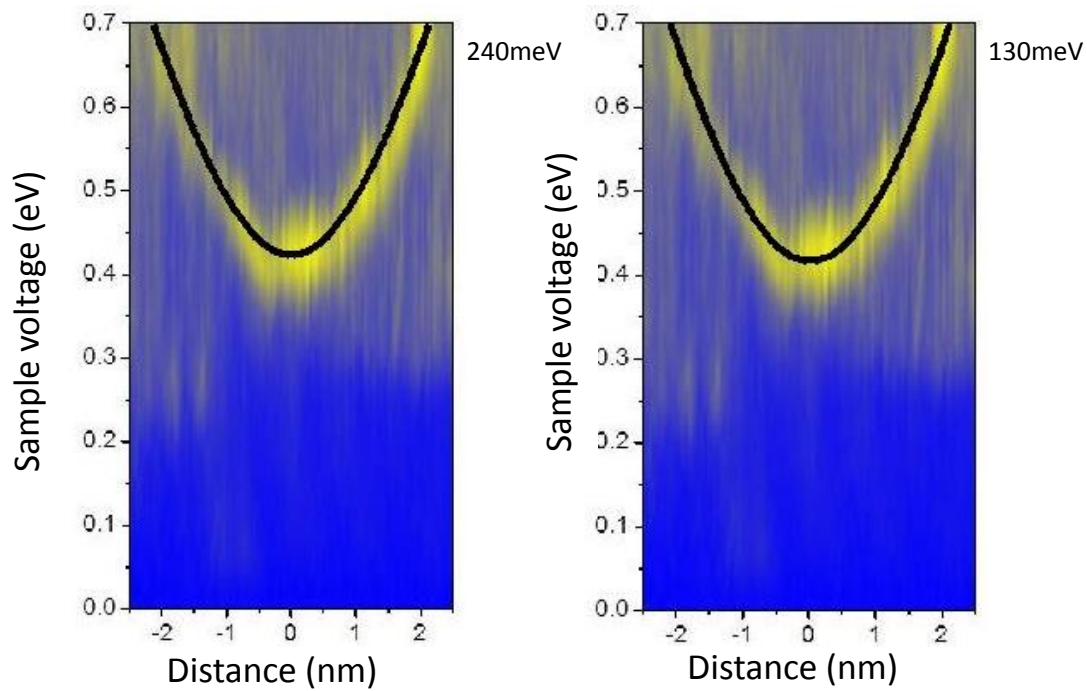